\documentclass{PoS}

\usepackage{epsfig}

\title{Charmed signatures for phase transitions in heavy-ion collisions}

\ShortTitle{Charmed signatures for phase transitions in heavy-ion
collisions}

\author{\speaker{Elena Bratkovskaya}%
\\
Frankfurt Institute for Advanced Studies, \\
 Ruth-Moufang-Str. 1, %
 60438 Frankfurt am Main, %
 Germany \\
         E-mail: \email{Elena.Bratkovskaya@th.physik.uni-frankfurt.de}}

\author{Olena Linnyk\\
Frankfurt Institute for Advanced Studies, \\
 Ruth-Moufang-Str. 1, %
 60438 Frankfurt am Main, %
 Germany
}

\author{Wolfgang Cassing\\
Institut f\"ur Theoretische Physik,
  Universit\"at Giessen,\\
  Heinrich--Buff--Ring 16, %
  35392 Giessen, %
  Germany
}

\author{Horst St\" ocker\\
Institut f\"ur Theoretische Physik, 
 Johann Wolfgang Goethe University,\\
 Max-von-Laue-Str. 1, %
 60438 Frankfurt am Main, %
 Germany \\
 Frankfurt Institute for Advanced Studies, \\
 Ruth-Moufang-Str. 1, %
 60438 Frankfurt am Main, %
 Germany
}

\abstract{The interplay of charmonium production and suppression
in $In+In$ and $Pb+Pb$ reactions at 158~AGeV and in $Au+Au$
reactions at $\sqrt{s}=200$~GeV is investigated with the HSD
transport approach within the `hadronic comover model' and the
`QGP melting scenario'. The results for the $J/\Psi$ suppression
and the $\Psi'$ to $J/\Psi$ ratio are compared to the recent data
of the NA50, NA60, and PHENIX Collaborations. We find that, at
158~AGeV, the comover absorption model performs better than the
scenario of abrupt threshold melting. However, neither interaction
with hadrons alone nor simple color screening satisfactory
describes the data at $\sqrt{s}=200$~GeV. A deconfined phase is
clearly reached at RHIC, but a theory having the relevant degrees
of freedom in this regime (strongly interacting quarks/gluons) is
needed to study its transport properties.}

\FullConference{Critical Point and Onset of Deconfinement   -  4th
International Workshop\\
         July 9 - 13, 2007\\
         Darmstadt, Germany}

\begin{document}

\section{Introduction}

Measurements of charmonium production in heavy-ion collisions at
different energies can provide clear signatures of the onset of
deconfinement. Indeed, according to potential model predictions 
and to the pioneering idea of Matsui and Satz~\cite{Satz},
$c\bar{c}$ meson states might no longer be formed in a very hot
fireball due to color screening~\cite{Satznew,Satzrev,KSatz}. This
initially intuitive expectation has guided experimental studies
for almost two decades. However, more recent lattice QCD
calculations have shown that the $J/\Psi$ survives up to at least
1.5 $T_c$ ($T_c \approx$ 170 to 185~MeV) such that the lowest
$c\bar{c}$ states may remain bound up to rather high energy
density~\cite{Aarts,KarschJP,HatsudaJP,Karsch2}. On the other
hand, the $\chi_c$ and $\Psi^\prime$ appear to melt soon above
$T_c$.


According to present knowledge, the charmonium production in
heavy-ion collisions, {\it i.e.}  $c\bar{c}$ pairs,  occurs
exclusively at the initial stage of the reaction in primary
nucleon-nucleon collisions. At the very early stage color dipole
states are expected to be formed ({\it cf.}
Refs.~\cite{Kharzeev,Capella}). These $c\bar{c}$ states are
assumed to be absorbed in a `pre-resonance state' before the final
hidden charm mesons are formed.  Such absorption -- denoted by
`normal nuclear suppression' -- is also present in $p+A$ reactions
and is determined by a dissociation cross section $\sigma_B$
$\sim$ 4 to 7 mb. Those charmonia or `pre-resonance' states  that
survive normal nuclear suppression during the short overlap phase
of the Lorentz contracted nuclei furthermore suffer from (i)  a
possible dissociation in the deconfined medium at sufficiently
high energy density and (ii) the interactions with secondary
hadrons (comovers) formed in a later stage of the nucleus-nucleus
collision.

In the QGP `threshold scenario', e.g the geometrical Glauber model
of Blaizot et al.~\cite{Blaizot} as well as the percolation model
of Satz~\cite{Satzrev}, the QGP suppression `(i)' sets in rather
abruptly as soon as the energy density exceeds a threshold value
$\varepsilon_c$, which is a free parameter. This version of the
standard approach  is motivated by the idea that the charmonium
dissociation rate is drastically larger in a quark-gluon-plasma
(QGP)  than in a hadronic medium~\cite{Satzrev}. On the other
hand, the extra suppression of charmonia in the high density phase
of nucleus-nucleus collisions at SPS
energies~\cite{NA50aa,NA50b,NA50a,NA60} has been attributed to
inelastic comover scattering ({\it
cf.}~\cite{Capella,Cass97,Cass99,Vogt99,Gersch,Cass00,Kahana,Spieles,Gerland}
and Refs. therein) assuming that the corresponding $J/\Psi$-hadron
cross sections are in the order of a few
mb~\cite{Haglin,Konew,Ko,Sascha}. In these models `comovers' are
viewed not as asymptotic hadronic states in vacuum but rather as
hadronic correlators (essentially of vector meson type) that might
well survive at energy densities above 1 GeV/fm$^3$. Additionally,
alternative absorption mechanisms  might play a role, such as
gluon scattering on color dipole states as suggested in
Refs.~\cite{Kojpsi,Rappnew,Blaschke1,Blaschke2} or charmonium
dissociation in the strong color fields of overlapping
strings~\cite{Geiss99}.

We recall that apart from absorption or dissociation channels for
charmonia also recombination channels such $D+ \bar{D} \rightarrow
X_c + meson$ ($X_c =(J/\Psi, \chi_c, \Psi^\prime)$) play a role in
the hadronic phase. These backward channels -- relative to
charmonium dissociation with comoving mesons -- have been found to
be practically negligible at the SPS energies~\cite{Olena}, but
extremely important at the top RHIC energy of
$\sqrt{s}=200$~GeV~\cite{Olena2}. This is in accordance with
independent studies in Refs.~\cite{Ko,Rappnew,PBM,PBM2} and
earlier analysis within the HSD transport
approach~\cite{brat03,brat04}.

The explicit treatment of initial $c\bar{c}$ production by primary
nucleon-nucleon collisions and the implementation of the comover
model - involving a single matrix element $M_0$ fixed by the data
at SPS energies - as well as the QGP threshold scenario in HSD
were explained in Ref.~\cite{Olena} (see Fig.~1 of
Ref.~\cite{Olena} for the relevant cross sections). We recall that
the `threshold scenario' for charmonium dissociation is
implemented as follows: whenever the local energy density
$\varepsilon(x)$ is above a threshold value $\varepsilon_j$ (where
the index $j$ stands for $J/\Psi, \chi_c, \Psi^\prime$), the
charmonium is fully dissociated to $c + \bar{c}$. The default
threshold energy densities adopted are $\varepsilon_1 = 16$
GeV/fm$^3$ for $J/\Psi$, $\varepsilon_2 = 2$ GeV/fm$^3$ for
$\chi_c$, and $\varepsilon_3 =2 $ GeV/fm$^3$ for $ \Psi^\prime$.

It is presently not clear, if also the $D$-mesons survive at
temperatures $T > T_c$, but strong correlations between a light
quark (antiquark) and a charm antiquark (quark) are likely to
persist~\cite{Rapp05}. One may also speculate that similar
correlations survive also in the light quark sector above $T_c$
such that `hadronic comovers' -- most likely with different
spectral functions -- might show up also at energy densities above
1~GeV/fm$^3$, which is taken as a characteristic scale for the
critical energy density. Therefore, we study both possibilities:
{\em with} and {\em without} comover absorption (and $D+\bar D$
recombination) at energy densities above the cut-energy density
parameter $\epsilon_{cut}=1$~GeV/fm$^3$.
\begin{figure}
\centerline{\psfig{figure=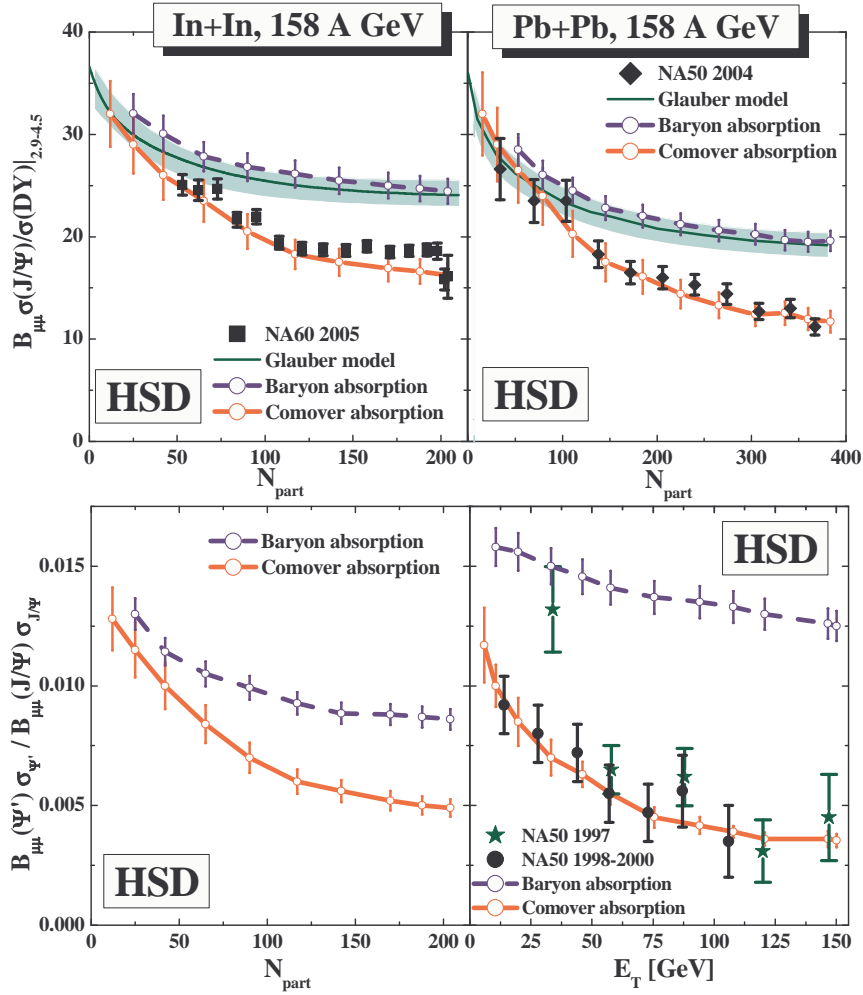,width=0.75\textwidth}}
\caption{The ratio $B_{\mu\mu}\sigma(J/\Psi) / \sigma(DY)$  as a
function of the number of participants in In+In (l.h.s.) and Pb+Pb
reactions (r.h.s.) at 158 A$\cdot$GeV. The full symbols denote the
data from the NA50 and NA60 Collaborations (from Refs.
\protect\cite{NA50O,NA60,NA50PsiPrime}), while the dashed (blue)
lines represent the HSD calculations including only dissociation
channels with nucleons. The lower parts of the figure show the HSD
results in the same limit for the $\Psi^\prime$ to $J/\Psi$ ratio
as a function of $N_{part}$ (for In+In) or the transverse energy
$E_T$ (for Pb+Pb). The solid (red) lines show the HSD results for
the comover absorption model with a matrix element squared
$|M_0|^2$ = 0.18 fm$^2$/GeV$^2$.  The (light blue) bands in the
upper parts of the figure give the estimate for the normal
 nuclear $J/\Psi$ absorption as calculated by the NA60 Collaboration.
 The vertical lines on
the graphs reflect the theoretical uncertainty due to limited
statistics of the calculations. The figure is taken
from~\cite{Olena}.} \label{set5}
\end{figure}

Since we aimed to answer, whether the charmonium dissociation
mechanism is identical at SPS and top RHIC energies, we adopted
in~\cite{Olena2} the same cross sections for the color-dipole
dissociation with nucleons as well the dissociation cross sections
with comovers as in Ref.~\cite{Olena} for SPS. Consequently no
free parameters entered our studies at the RHIC energy. We note
that the hadronic comover reactions for the recreation of
charmonia $J/\Psi, \chi_c, \Psi^\prime$ by $D + \bar{D}$ reactions
are incorporated in all simulations. This is a `default' in the
comover absorption and recreation scenario and `necessary' in the
QGP `threshold scenario' because (in view of
Fig.~\ref{RHICthreshold}, l.h.s.) practically all charmonia are
dissolved due to the very high initial energy densities.
Therefore, any model without recreation of charmonia is clearly
ruled out by the PHENIX data.
\begin{figure*}[!]
\centerline{\psfig{figure=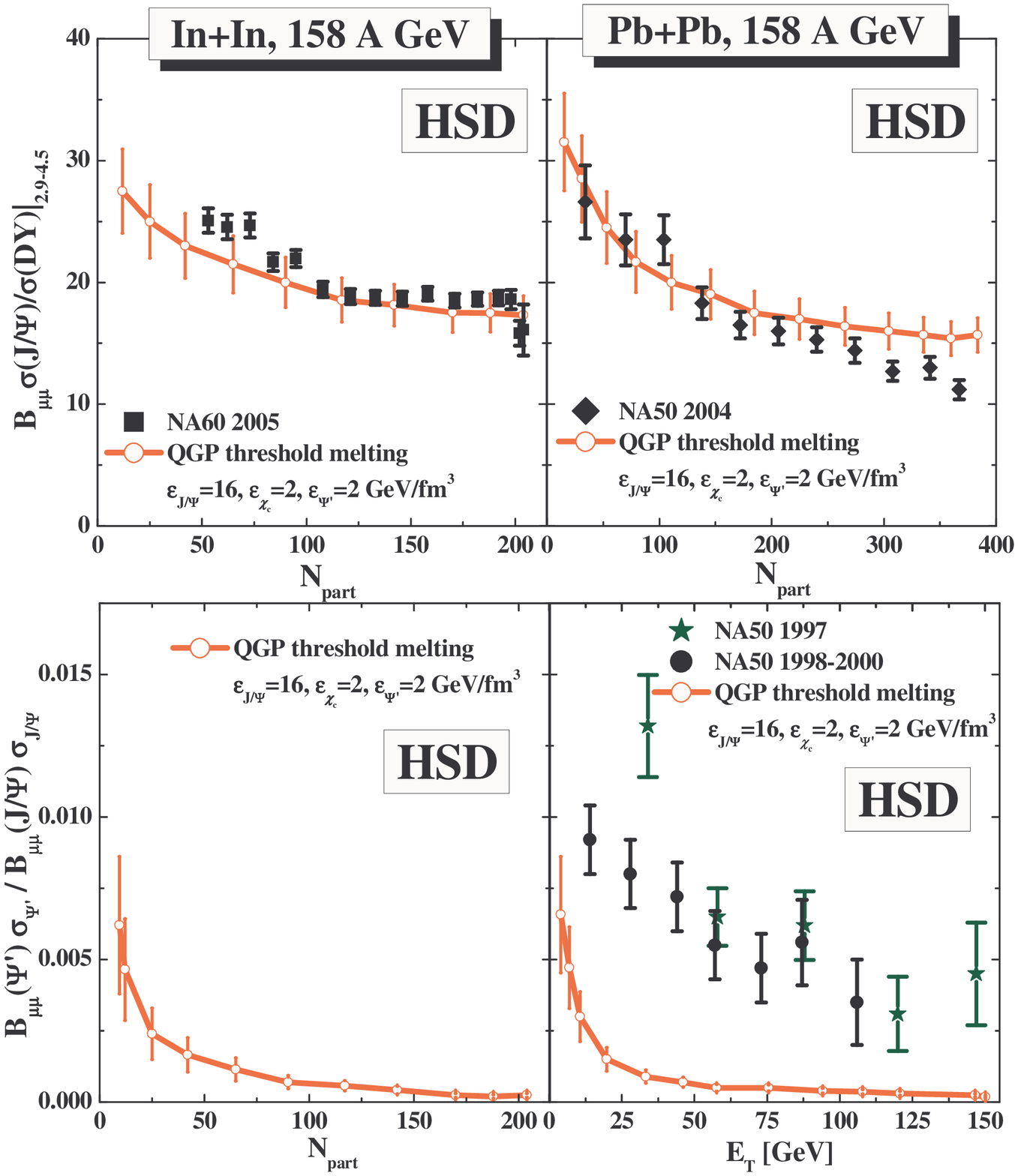,width=0.75\textwidth}}
\caption{Same as Fig. \protect\ref{set5} but for the `QGP
threshold scenario' with $\varepsilon_{J/\Psi} = 16$ GeV/fm$^3$,
$\varepsilon_{\chi_c} = 2$ GeV/fm$^3$ =
$\varepsilon_{\Psi^\prime}$ while discarding comover absorption.
The figure is taken from~\cite{Olena}.} \label{set9}
\end{figure*}

\section{Comparison to data}

We directly step on with results for the charmonium suppression at
SPS energies in comparison with the experimental data from the
NA50 and NA60 Collaborations.  These Collaborations present their
results on $J/\Psi$ suppression as the ratio of the dimuon decay
of $J/\Psi$ relative to the Drell-Yan background from 2.9 - 4.5
GeV invariant mass as a function of the transverse energy $E_T$,
or alternative, as a function of the number of participants
$N_{{\rm part}}$, {\it i.e.}
\begin{equation} \label{rat} B_{\mu\mu}\sigma(J/\Psi) /
\sigma(DY)|_{2.9-4.5},
\end{equation}
where $B_{\mu\mu}$ is the branching ratio for $J/\Psi\to
\mu^+\mu^-$. In order to compare our calculated results to
experimental data, we need an extra input, i.e. the normalization
factor $B_{\mu\mu}\sigma_{NN}(J/\Psi) / \sigma_{NN}(DY)$, which
defines the $J/\Psi$ over Drell-Yan ratio for elementary
nucleon-nucleon collisions. We choose
$B_{\mu\mu}\sigma_{NN}(J/\Psi) / \sigma_{NN}(DY) = 36$ in line
with the NA60 compilation~\cite{NA60}.
\begin{figure*}
\centerline{\psfig{figure=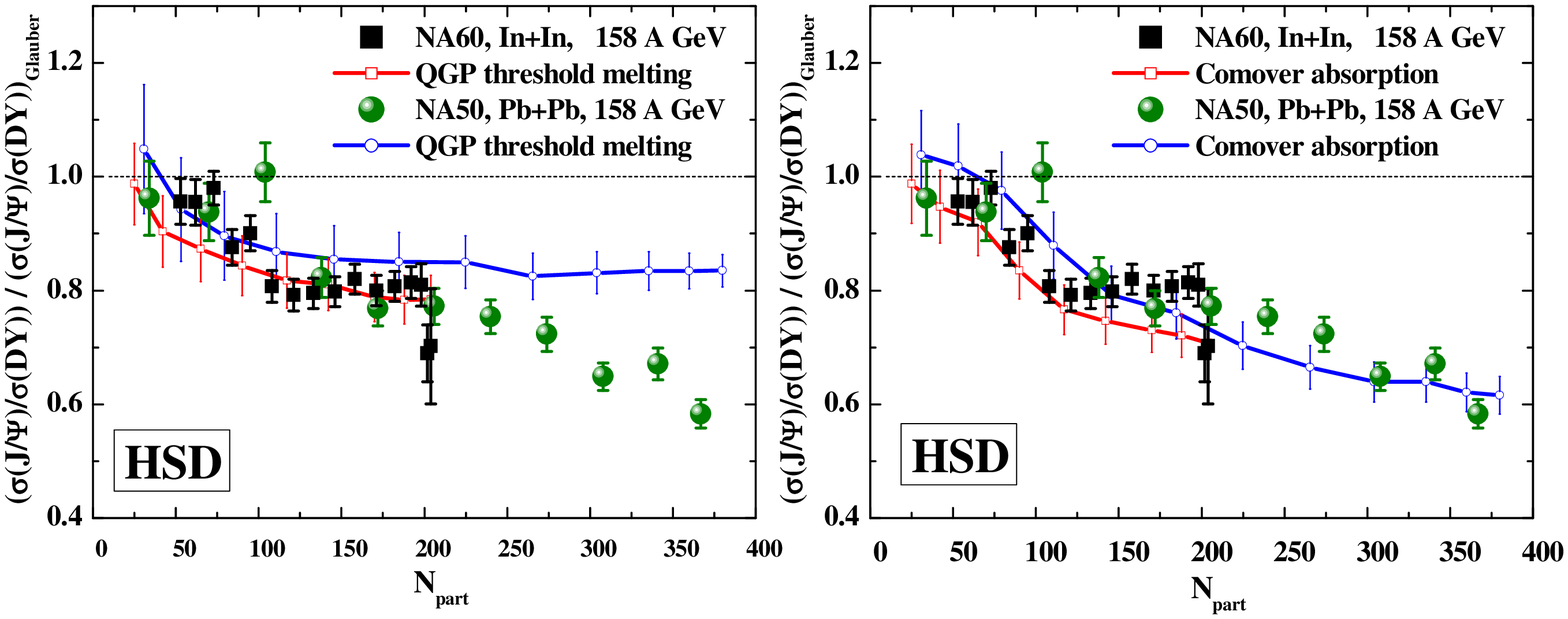,width=0.98\textwidth}}
\caption{The ratio $B_{\mu\mu}\sigma(J/\Psi) / \sigma(DY)$  as a
function of the number of participants $N_{part}$ in In+In (red
line with open squares) and Pb+Pb reactions (blue line with open
circles) at 158 A$\cdot$GeV relative to the normal nuclear
absorption given by the straight black line. The full dots and
squares denote the respective data from the NA50 and NA60
Collaborations. The model calculations reflect the comover
absorption model (right part) and the `QGP threshold scenario'
(left part) with $\varepsilon_{J/\Psi} = 16$ GeV/fm$^3$,
$\varepsilon_{\chi_c} = 2$ GeV/fm$^3$, $\varepsilon_{\Psi^\prime}$
= 6.55 GeV/fm$^3$ while discarding comover absorption. Figure is
taken from~\cite{Olena}.} \label{ratioJP}
\end{figure*}

Furthermore, the $\Psi^\prime$ suppression is presented
experimentally by the ratio
\begin{equation} \label{pis}
\frac{B_{\mu\mu}(\Psi^\prime\to
\mu\mu)\sigma(\Psi^\prime)/\sigma(DY) } {B_{\mu\mu}(J/\Psi\to
\mu\mu)\sigma(J/\Psi) / \sigma(DY)}.
\end{equation}
In our calculations we adopt this ratio to be 0.0165 for
nucleon-nucleon collisions, which is again based on the average
over $pp, pd, pA$ reactions~\cite{NA50_03}.

We first show in Fig. \ref{set5} the calculated ratio (\ref{rat})
as a function of $N_{{\rm part}}$ for Pb+Pb and In+In collisions
at 158 A$\cdot$GeV (upper plots) in the nuclear suppression
scenario, i.e. without comover dissociation or `QGP threshold
suppression'. The dashed (blue) lines stand for the HSD result
while the (light blue) bands  give the estimate for the normal
nuclear $J/\Psi$ absorption as calculated by the NA60
Collaboration. The normal nuclear suppression from HSD is seen to
be slightly lower than the (model dependent) estimate from NA60,
however, agrees quite well with their model calculations for more
central reactions. The various experimental data points have been
taken from Refs.~\cite{NA60,NA50O,NA50PsiPrime}.
%
As a next step we add the comover dissociation channels within the
model described in~\cite{Olena} for a matrix element squared
$|M_0|^2$ = 0.18 fm$^2$/GeV$^2$. Note that in this case the
charmonium reformation channels are incorporated, too, but could
be discarded since the charmonium regeneration is negligible at
SPS energies (cf. Ref.~\cite{brat03}). The extra suppression of
charmonia by comovers is seen in Fig.~\ref{set5} (solid red lines)
to match the $J/\Psi$ suppression in In+In and Pb+Pb as well as
the $\Psi^\prime$ to $J/\Psi$ ratio (for Pb+Pb) rather well. The
more recent data (1998-2000) for the $\Psi^\prime$ to $J/\Psi$
ratio agree with the HSD prediction within error bars. This had
been a problem in the past when comparing to the 1997 data (dark
green stars). The $\Psi^\prime$ to $J/\Psi$ ratio for In+In versus
centrality is not yet available from the experimental side but the
theoretical predictions are provided in Fig.~\ref{set5} and might
be approved/falsified in near future.
\begin{figure}
\centerline{ \psfig{figure=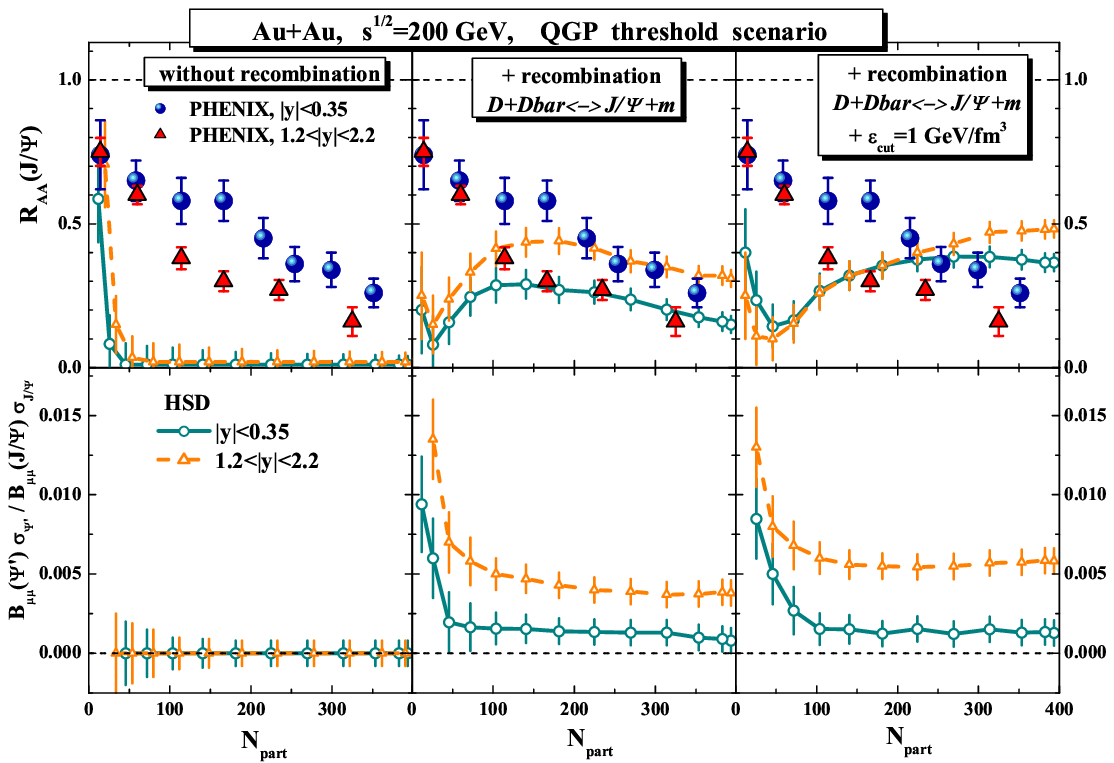,width=\columnwidth}}
\caption{The $J/\Psi$ nuclear modification factor $R_{AA}$ for
$Au+Au$ collisions at $\sqrt{s}=200$~GeV as a function of the
number of participants $N_{part}$ in comparison to the data from
[10] for midrapidity (full circles) and forward rapidity (full
triangles). HSD results for the QGP `threshold melting' scenarios
are displayed in terms of the lower (green solid) lines for
midrapidity $J/\Psi$'s ($|y| \le 0.35$) and in terms of the upper
(orange dashed) lines for forward rapidity ($1.2 \le y \le 2.2$)
within different recombination scenarios (see text). The error
bars on the theoretical results indicate the statistical
uncertainty due to the finite number of events in the HSD
calculations. Predictions for the ratio $B_{\mu \mu} (\Psi ')
\sigma _{\Psi'} / B_{\mu \mu} (J/\Psi) \sigma _{J/\Psi} $ as a
function of the number of participants $N_{part}$ are shown in the
lower set of plots. The figure is taken from~\cite{Olena2}.}
\label{RHICthreshold}
\end{figure}

The results for the `threshold scenario' are displayed in
Fig.~\ref{set9} in comparison to the same data for the thresholds
$\varepsilon_{J/\Psi} = 16$ GeV/fm$^3$, $\varepsilon_{\chi_c} = 2$
GeV/fm$^3$ = $\varepsilon_{\Psi^\prime}$. In this scenario the
$J/\Psi$ suppression is well described for In+In but the
suppression is slightly too weak for very central Pb+Pb reactions.
This result emerges since practically all $\chi_c$ and
$\Psi^\prime$ dissolve for $N_{part}
>$ 100 in both systems whereas the $J/\Psi$ itself survives at the
energy densities reached in the collision. Since the nucleon
dissociation is a flat function of $N_{part} $ for central
reactions, the total absorption strength is flat, too. The
deviations seen in Fig.~\ref{set9} might indicate a partial
melting of the $J/\Psi$ for $N_{part} > $ 250, which is not in
line with most lattice QCD calculations claiming at least
$\varepsilon_{J/\Psi}>$ 5 GeV/fm$^3$. In fact, a lower threshold
of  5 GeV/fm$^3$ (instead of 16 GeV/fm$^3$) for the $J/\Psi$ has
practically no effect on the results shown in Fig.~\ref{set9}.
Furthermore, a threshold energy density of 2 GeV/fm$^3$ for the
$\Psi^\prime$ leads to a dramatic reduction of the $\Psi^\prime$
to $J/\Psi$ ratio which is in severe conflict with the data (lower
part of Fig.~\ref{set9}). Also note that there is no step in the
suppression of $J/\Psi$ versus centrality. As pointed out before
by Gorenstein et al. in Ref.~\cite{Goren}, this is due to energy
density fluctuations in reactions with fixed $N_{part}$ (or
$E_T$).
\begin{figure}
\centerline{\psfig{figure=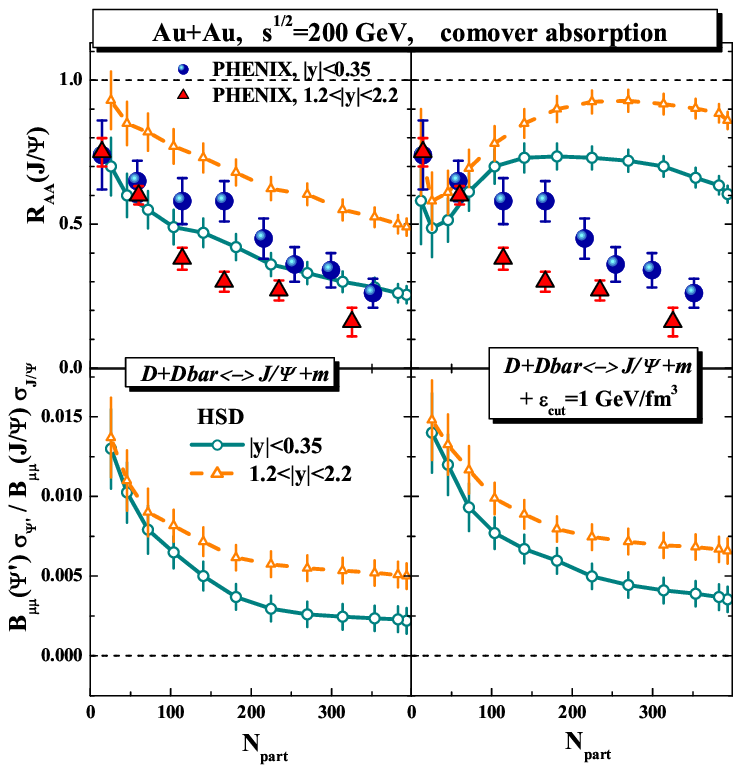,width=0.7\columnwidth}}
\caption{Same as Fig.~4 for the `comover absorption scenario'
including the charmonium reformation channels without cut in the
energy density (l.h.s.) and with a cut in the energy density
$\epsilon _{cut} = 1$~GeV/fm$^3$ (see text for details). The
figure is taken from~\cite{Olena2}. } \label{RHICcomover}
\end{figure}

Additionally, one can plot the results in an intuitive though
model-dependent way, as a ratio of the measured $J/\Psi$ yield
divided by the normal nuclear absorption result calculated in the
Glauber model. Since the NA60 Collaboration prefers to represent
their data in this form, we additionally show in Fig.
\ref{ratioJP} our calculations for In+In (red lines with open
squares) and Pb+Pb (blue lines with open circles) as a function of
the number of participants $N_{part}$ relative to the normal
nuclear absorption given by the straight black line\footnote{Note
that recently the NA60 collaboration has refitted the parameters
of their Galuber model, therefore newer data
releases~\cite{NA60newrelease} might appear to be up- or
down-scaled compared to the data plotted here~\cite{NA60}, if
shown in this particular representation (measured to expected
ratio). This scaling falls within the systematic uncertainty of
the ratio and does not change results and conclusions of our
study.}. The full dots and squares denote the respective data from
the NA50 and NA60 Collaborations. The model calculations reflect
the comover absorption model (right part) and the `QGP threshold
scenario' (left part) with $\varepsilon_{J/\Psi} = 16$~GeV/fm$^3$,
$\varepsilon_{\chi_c} = 2$~GeV/fm$^3$, $\varepsilon_{\Psi^\prime}
= 6.55$~GeV/fm$^3$. Since only the representation is different the
message stays the same: The comover absorption model follows
slightly better the fall of the $J/\Psi$ survival probability with
increasing centrality whereas the `threshold scenario' leads to an
approximate plateau in both reactions for high centrality.

Let us now move to a much higher energy scale by calculating
charmonium dynamics at the top RHIC energy of $\sqrt{s}=200$~GeV.
In the initial stages of $Au+Au$ collisions at this $\sqrt{s}$,
energy densities above 30~GeV/fm$^3$ are reached~\cite{Olena2}.
Therefore, in the threshold melting scenario, all initially
created $J/\Psi$, $\Psi'$ and $\chi _c$ mesons melt. However, the
PHENIX collaboration has found that at least 20\%  of $J/\Psi$ do
survive at RHIC~\cite{PHENIX}. Thus, the importance of charmonium
recreation is shown again. We account for $J/\Psi$ recreation via
the $D  \bar D$ annihilation processes as explained in detail
in~\cite{Olena,Olena2}. Note that in our approach, the cross
sections of charmonium recreation in $D + \bar D \to J/\Psi +
meson$ processes is fixed by detailed balance from the comover
absorption cross section $J/\Psi + meson \to D + \bar D$. But even
after both these processes are added to the threshold melting
mechanism, the centrality dependence of the $R_{AA} (J/\Psi)$
cannot be reproduced, especially in the peripheral collisions (see
Fig.~\ref{RHICthreshold}). This holds for both possibilities: with
(r.h.s. of Fig.~\ref{RHICthreshold}) and without (center of
Fig.~\ref{RHICthreshold}) the energy density cut $\epsilon_{cut}$,
below which $D$-mesons and comovers exist and can participate in
$D + \bar D \leftrightarrow J/\Psi + meson$ reactions.

We recall that  the nuclear modification factor $R_{AA}$ is given
by
\begin{equation}
R_{AA}=\frac{ d N (J/\Psi) _{AA} / d y  }{ N_{coll} \cdot d N
(J/\Psi) _{pp} / d y },
\end{equation}
where $d N (J/\Psi) _{AA} / d y $ denotes the final yield of
$J/\Psi$ in $A A$ collisions, $d N (J/\Psi) _{pp} / d y$ is the
yield in elementary $p p$ reactions, $N_{coll}$ is the number of
binary collisions.

Comover absorption scenarios give generally a correct dependence
of the yield on the centrality. If an existence of D-mesons at
energy densities above 1 GeV/fm$^3$ is assumed, the amplitude of
suppression of $J/\Psi$ at mid-rapidity is also well reproduced
(see the line for `comover without $\epsilon_{cut}$' scenario in
Fig.\ref{RHICcomover}, l.h.s.). Note that this line correspond to
the prediction made in the HSD approach in~\cite{brat04}. 
On the other hand, the rapidity dependence of the comover result
is wrong, both with and without $\epsilon _{cut}$. If hadronic
correlators exist only at $\epsilon < \epsilon _{cut}$, comover
absorption is insufficient to reproduce the $J/\Psi$ suppression
even at mid-rapidity (see Fig.~\ref{RHICcomover}, r.h.s.). The
difference between the theoretical curves marked `comover +
$\epsilon _{cut}$' and the data shows the maximum supression that
can be attributed to a deconfined medium.

%
%
%

\section{Summary}

We have investigated the formation and suppression dynamics of
$J/\Psi$, $\chi_c$ and $\Psi^\prime$ mesons within the HSD
transport approach for $In+In$ and $Pb+Pb$ reactions at 158~AGeV
and for $Au+Au$ reactions at $\sqrt{s}$ = 200 GeV. Two currently
discussed models, i.e. the 'hadronic comover absorption and
reformation' model as well as the 'QGP threshold melting scenario'
have been compared to the available experimental data. We adopted
the same parameters for cross sections (matrix elements) or
threshold energies at both bombarding energies.

We find that both scenarios are compatible with experimental
observation of $J/\Psi$ suppression at SPS energies, while the
$\Psi'$ to $J/\Psi$ ratio data appear to be in conflict with the
`threshold melting' scenario~\cite{Olena}. On the other hand, both
`comover absorption' and `threshold melting' fail severely at RHIC
energies~\cite{Olena2}. The failure of the 'hadronic comover
absorption' model goes in line with its underestimation of the
collective flow $v_2$ of leptons from open charm decay as
investigated in Ref.~\cite{brat05}. This suggests that 1) a
deconfined phase is clearly reached at RHIC, 2) the dynamics of
$c, \bar{c}$ quarks at this energy are dominated by partonic
interactions in the strong QGP (sQGP) which cannot be modeled by
`hadronic' interactions or described appropriately by color
screening alone.




\bibliographystyle{h-physrev3}
\bibliography{HSDcharm}

\end{document}